\documentclass[times,authoryear]{paper}
\usepackage{framed,multirow}
\usepackage{graphicx}

\usepackage{amssymb}
\usepackage{latexsym}

\usepackage{url}
\usepackage{xcolor}
\definecolor{newcolor}{rgb}{.8,.349,.1}

\usepackage[citebordercolor=white]{hyperref}

\begin{document}

%% \verso{Oleg Berngardt \textit{etal}}

%% \begin{frontmatter}

\title{Self-learning signal classifier for decameter coherent scatter radars} %\tnoteref{tnote1}}%
%%% \tnotetext[tnote1]{This is an example for title footnote coding.}

\author{Oleg Berngardt \and Ivan Lavygin \and et al.}

\maketitle

\begin{abstract}
%%%
The paper presents a method for automatic constructing a classifier
for processed data obtained by decameter coherent scatter radars.
Method is based only on the radar data obtained, the results
of automatic modeling of radio wave propagation in the ionosphere,
and mathematical criteria for estimating the quality of the models.
The final classifier is the model trained 
at data obtained by 12 radars of the SuperDARN and SECIRA networks 
over two years for each radar. 
The number of the model coefficients is 2669. For the
classification, the model uses both the calculated parameters of radio
wave propagation in the model ionosphere and the parameters directly
measured by the radar. Calibration of radiowave elevation measurements at each
radar was made using meteor trail scattered signals. The analysis showed
that the optimal number of classes in the data is 37, of which 25
are frequently observed. The analysis made it possible to choose 
14 classes from them, which are confidently separated in other variants
of model training. A preliminary interpretation of 10 of them was
carried out. The dynamics of observation of various classes and their
dependence on the geographical latitude of radars at different levels
of solar and geomagnetic activity were presented, it was shown that
it does not contradict with known physical mechanisms. The analysis
showed that the most important parameters to identify the classes
are the shape of the signal ray-tracing trajectory in its
second half, the ray-traced scattering height and
the Doppler velocity measured by the radar.
%%%%
\end{abstract}

%% \begin{keyword}
%% Keywords
%% \KWD radar data processing\sep neural networks\sep automatic classification\sep data-driven analysis\sep SECIRA\sep SuperDARN
%% \end{keyword}

%% \end{frontmatter}

%% For linenumbers

\section{Introduction}

Decameter coherent scatter radars are currently popular and highly
informative instruments for monitoring the state of the ionosphere
at high and mid-latitudes. The largest network of these radars today is SuperDARN
- Super Dual Auroral Radar Network, \cite{Greenwald_1995c,Chisham_2007c,Nishitani_2019}.
There are also CN-DARN\cite{Zhang_2024} and SECIRA\cite{Berngardt_2020_istp}
networks growing. The features of the radars are their short-wave radio frequency range,
allowing long-range (multihop) propagation of radio waves, significantly
expanding the radar range upto 3-4 thousand kilometers and
increasing the number of effects affecting the radiowave propagation,
but complicating the interpretation of the received signals\cite{Nishitani_2019}.

The first stage of interpreting the signals is their classification
- separating them into classes based on their characteristics. The one
of the oldest and simplest methods of the classification is separating them
into ionospheric scattering and scattering from the earth's surface
(groundscatter), traditionally carried out by the spectral width
of the received signal and its Doppler velocity \cite{Ponomarenko_2007}.
Due to the large number of different signal types\cite{Nishitani_2019}, 
the classification is not complete and requires improvement.
Recent approaches include the use of statistical methods for separating the two classes \cite{Blanchard_2009,Lavygin_2019},
the use of clustering for taking into account the spatio-temporal
distribution of the signals \cite{Ribeiro_2011}, the use of neural
networks and clustering methods to identify complex dependencies and classes \cite{Kunduri_2022,KONG_2024}.
Most of these approaches are based on separating the signals into initially predefined classes.

Another approach to solving this problem is proposed in \cite{Berngardt_2022a}
and developed in \cite{Berngardt_2022b,Berngardt_2025} - 
a method for separating signals into initially unknown classes.
It is performed within the framework of self-learning neural networks -
the superposition of unsupervised and supervised machine learning
methods, with subsequent interpretation of the resulting classes by
researcher after final network train. This approach refers to data-driven
analysis methods and is based on the use of the experimental data
only to build an optimal classification model automatically, by reducing researcher influence.

Previously, these works were carried out only for the SECIRA network
radars\cite{Berngardt_2022a,Berngardt_2022b,Berngardt_2025}. 
The proposed paper is a generalization of this approach.
We attempt to further reduce the influence of the researcher on the
process of constructing the model and to generalize the model to the
periods of different solar activity and different radar locations.

We also study the dependence of observed signal classes on geophysical 
conditions and radar location, as well as importance of different measured 
signal parameters on classification accuracy.

\begin{figure}
\centering
\includegraphics[scale=0.25]{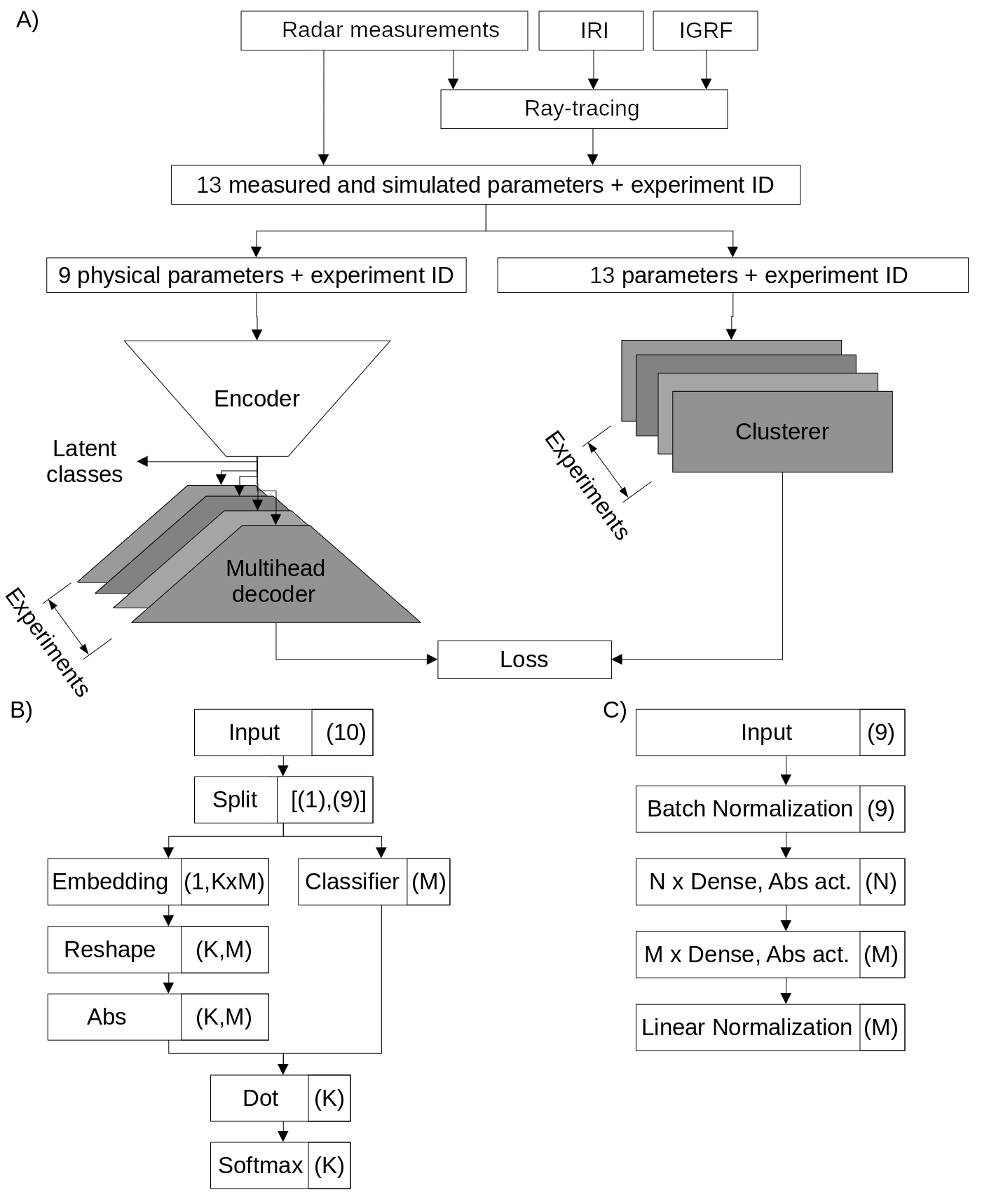}
\caption{A) Neural network architecture and its training method. 
Different colors corresponds to different experiments. Number of decoder heads 
and clusterers is about 4000; 
B) Implementation of a multi-head autoencoder; 
C) Implementation of the Encoder (classifier).}
\label{fig:architect}
\end{figure}

The proposed classifier model (Encoder in Fig.\ref{fig:architect}A)
is a neural network of an extremely simple architecture (two fully
connected layers), sufficient to solve the problem. Its architecture
is shown in Fig.\ref{fig:architect}C. The main difficulty of this approach is the
process of classifier training, which includes building a huge model (multi-head
autoencoder) based on it, each head of which is trained independently
in independent experiments. The network teacher (labeler) in this case is a clusterer
- a method for separating signals into groups (clusters) without a teacher. This
approach was proposed in \cite{Berngardt_2022a} and was referred
as 'wrapped classifier' to highlight the fact that to train a classifier,
it is necessary to train a more complex neural network ('wrap'),
of which this classifier will be only a small part.

The architecture of autoencoder networks is quite popular and allows
extracting internal (latent) representations from data - the implicit
(hidden) independent parameters that describe the studied data well
\cite{Autoencoders}. As shows our analysis this architecture has demonstrated
its effectiveness for solving our classification problem. A detailed architecture of
the used autoencoder is shown in Fig.\ref{fig:architect}B. The autoencoder
is a significantly optimized and reduced version of the models \cite{Berngardt_2022a,Berngardt_2022b},
proposed and described in detail in \cite{Berngardt_2025}.

\section{Preliminary data preparation}
\label{sec1}
\subsection{Initial data and training}
Coherent scatter radars are over-the-horizon radars\cite{Greenwald_1995c}, that transmit pulse sounding sequences 
at frequencies in range 8-20MHz, and estimate the parameters of backscattered signals, as a function of 
time, delay(radar range) and transmit/receive azimuth. The parameters are estimated by accumulating autocorrelation 
function (ACF) of the signals received by two spatially separated antenna arrays to measure elevation angle of radiowave too. 
The measured ACFs are averaged over 2-6 seconds and processed by FITACF algorithm\cite{Ribeiro_2013}.
Each signal (measurement), measured at given time, range and azimuth has its own power and spectral characteristics, 
as well as elevation angle of the trajectory of received radiowave. These parameters and their spatio-temporal distribution 
are used later by researchers for the signal interpretation.
 
The following measured and model parameters were used to construct the neural
network classifier of the signals.

The parameters directly measured by the radar are:

1) Universal time and radar range;

2) Measured doppler velocity V and spectral width W, calculated from
the received signals using the FITACF algorithm \cite{Ribeiro_2013},
in the paper we use spectral width in the exponential model of the
correlation function (Wl);

3) Measured elevation angle of the received radiowave - determined
by the algorithm \cite{BERNGARDT_2021} after calibrating the radar
using meteor trail scattered signals;

4) Effective scattering height - calculated based on the elevation
angle and range as a result of refraction-free radio wave propagation.

The parameters obtained as a result of radio wave propagation simulation 
using direct radar measurments and the geometric optics method (ray-tracing) 
in the IRI-2020\cite{IRI_2020} model ionosphere are:

5) Trajectory elevation relative to the horizon at 4 points over the
range (1/4,2/4,3/4,4/4 of the measured range, we use sine of the elevation
angle);

6) Angle between the direction of radio wave trajectory and the Earth's
magnetic field at the scattering point  - calculated
from the reference magnetic field model IGRF\cite{IGRF}, we use cosine of the angle;

7) Propagation mode - the number of reflections (hops) from the underlying
layer (the earth's surface or the ionospheric layer in the case of
interlayer waveguide propagation) during propagation to the scatterer;

8) Scattering height.

~

Only 9 of these parameters were used as input data for the classifier:
velocity and spectral width, trajectory shape, angle with the magnetic
field, propagation mode, and scattering height. Whole 13 parameters
were used for clustering only. To train the model we separated all available 
data into experiments.
Each experiment corresponds to the data for a selected radar,
day, channel, azimuth (beam) of transmission, and
sounding frequency (with an accuracy of 1 MHz). To reduce the
amount of data and speed up calculations, only 3\% of all available
experiments were randomly selected. To limit the amount of the data,
no more than 5000 measurements (signals with a power of 3 dB above the noise
level) were randomly selected from each experiment. This gave us about 24 mln. unqiue unlabeled measurements 
in 5566 unique experiments for training the classification model.

The clusterer used is a modeling of the single experiment data distribution
by a mixture of multidimensional Gaussian distributions (Gaussian
Mixture)\cite{GMM} with the number of clusters detemined from the minimum
of the Bayesian information criterion (BIC)\cite{BIC}, each experiment has 
its own number of found clusters.

As a classifier we use two-layer neural network with absolute activation
functions, choosen, among other things, to provide the continuity with existing
generally used approach of separating the signals into two classes
\cite{Ponomarenko_2007} and to follow the Kolmogorov-Arnold representation theorem \cite{Kolmogorov_1957} 
allowing approximating a wide class of solutions by using two-layer networks only.
Linear normalization of the sum was used as output balancing 
for interpreting the results as class probabilities. The choice of
classifier and clusterer architecture is discussed in more detail
in \cite{Berngardt_2025}.

\subsection{Radars and their elevation calibration}

The methods that take into account radio
wave propagation requires accurate elevation angle measurements, and their
precise calibration.
The elevation calibration in our case was made using signals scattered
by meteor trails. The method was proposed for SuperDARN radars 
in \cite{CHISHAM_2013_meteors}. We use a modification of
the algorithm described in \cite{BERNGARDT_2021}, and based on calibrating the radar at each sounding frequency separately,
determining the scattering height for each meteor, and analyzing the expected and measured phase 
between signals received by two spatially separated phased arrays of the radar, 
calculated by FITACF algorithm.
In this paper, a small change was made to the algorithm: all meteor trails
were assumed to be observed at 104 km height. This is related with large
amount of calculations (data of all available radars over two years,
with sounding frequency descreet  200 kHz for calibration). 
So only averaged (processed) data, with significantly smaller amount than non-averaged 
IQ components of signals, can be used for calibration.
This does not allow direct use of the \cite{BERNGARDT_2021} method
for determining the scattering altitude by the meteor trail lifetime, 
that could be calculated by IQ components of the signal only.
The disadvantage of this modification is systematic
errors associated with neglected deviation of the scattering height from actual one, 
and the advantage is the calculation speed. In this case, we also assume
that the scattering altitudes do not depend on the geographical location
and time, which can also affects to the calibration accuracy. Nevertheless,
these assumptions allow us to unify the calibration of all the radars.

\begin{figure}
  \centering
  \includegraphics[scale=0.7]{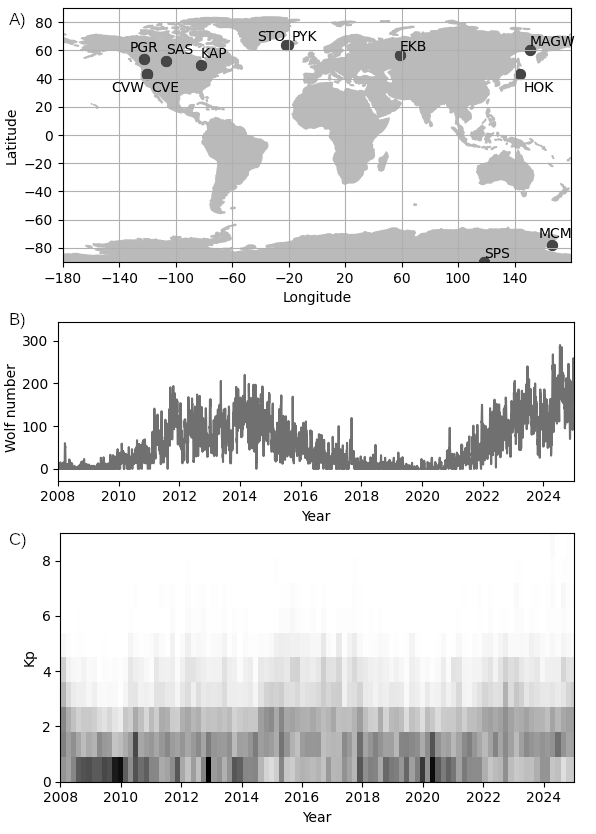}
  \caption{A) Radars used in the paper; B) Wolf numbers for the period 2008-2024 \cite{WolfNumbers}; C) Kp index distribution for 2008-2024 \cite{KpIndex}
}
\label{fig:map}
\end{figure}

During calibration, we select the signals detected by the FITACF algorithm\cite{Ribeiro_2013},
with low Doppler velocities (less than 50 m/s), high signal-to-noise
ratios (above 3dB), and from near radar ranges (not exceeding 350 km) 
were considered as scattering by meteor trails. All the radars
available for the analysis were calibrated, of which 12 radars were
selected. The radars are summarized in Table
\ref{tab:radyears}, their locations are shown in Fig.\ref{fig:map}A.
Most of the radars are from the northern hemisphere.
The geographic latitudes of the radars cover the range of 43-90 degrees
- from mid- to polar latitudes and cover both hemispheres. For each
radar we use the data over two years: one year of relatively low (2010,
2011, 2020, and 2021), and another year for relatively high solar activity
(2014, 2023, and 2024). Solar activity is shown in Fig.\ref{fig:map}B
by Wolf numbers. For any multi-channel radars except EKB
and MAGW, only one of the channels was analyzed, that provides the best
quality of elevation calibration and higher data amount. 

\begin{table}
\centering
\caption{Radars, their latitudes, and data years used in the research}
\begin{tabular}{|c|c|c|c|c|c|}
\hline 
\textbf{Radar} & \textbf{Lat.} & \textbf{Years} & \textbf{Radar} & \textbf{Lat.} & \textbf{Years}\tabularnewline
\hline 
\hline 
CVE SuperDARN & 43.3N & 2020,2023 & CVW SuperDARN & 43.3N & 2020,2023\tabularnewline
\hline 
EKB SECIRA & 56.5N & 2021,2023 & HOK SuperDARN  & 43.5N & 2020,2023\tabularnewline
\hline 
KAP SuperDARN  & 49.4N & 2020,2023 & MAGW SECIRA  & 60.0N & 2021,2024\tabularnewline
\hline 
MCM SuperDARN  & 77.8S & 2014,2020 & PGR SuperDARN & 54.0N & 2020,2023\tabularnewline
\hline 
PYK SuperDARN & 63.8N & 2010,2014 & SAS SuperDARN & 52.2N & 2020,2023\tabularnewline
\hline 
SPS SuperDARN & 90.0S & 2014,2021 & STO SuperDARN & 63.9N & 2011,2014\tabularnewline
\hline 
\end{tabular}\label{tab:radyears}
\end{table}

\begin{figure}
  \centering
  \includegraphics[scale=0.7]{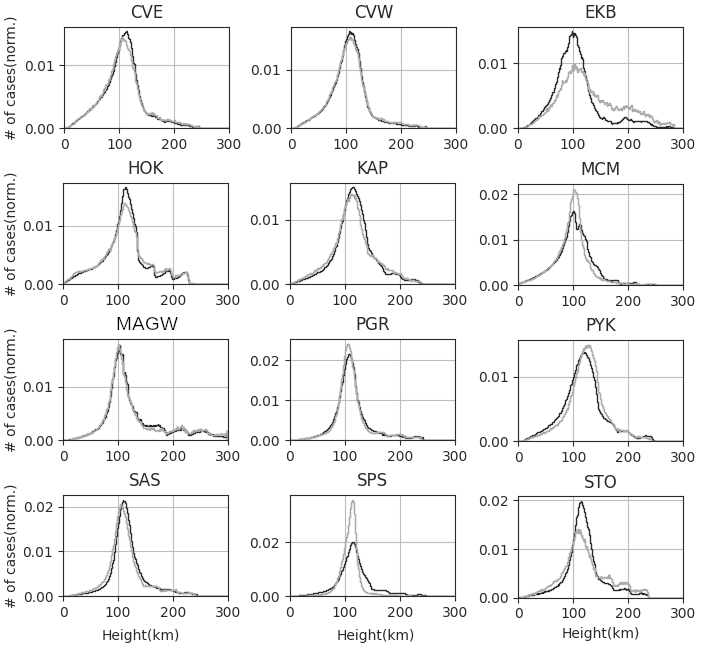}
  \caption{Quality of radar elevation calibration: meteor scattering altitude distributions (range $<$ 350km, Doppler drift velocity $<$ 50m/s) for each radar over the year. Black line - year of low solar activity, gray - year of high solar activity.}
  \label{fig:caliber}
\end{figure}

The scattering signal distributions at radar ranges up to 350 km after the elevation calibration
(usually interpreted as the scattering at altitudes of 90-100 km -
meteor echo\cite{Hall_1997} and near-range echo at E-layer altitudes\cite{Ponomarenko_2016nre})
over the altitudes for each radar and year are shown in Fig.\ref{fig:caliber}. One can see
from the figure that the elevation angle calibration is sufficiently
good, both for years of low and high solar activity: the most probable
scattering height corresponds to 100-110 km, well corresponding
to the expected types of scattering. We interpret the signals scattered
at altitudes of 150 km and above, for example by EKB radar, with a
large proportion of signals from the back lobe of the antenna pattern\cite{Milan_1997, Berngardt_2025b},
which is not taken into account during further analysis.

\section{Training the classifier}

\subsection{Selecting measurements for training}

An important problem in massive automatic signal processing is the presence
of signals comming not from the main lobe of the antenna pattern,
but from the back lobe\cite{Milan_1997}. The problem of excluding
such signals is quite complex. Their presence significantly distorts
the operation of the algorithm, since it requires accurate raytracing and calculating trajectory parameters.

The selection of main-beam signals for training the
model was carried out using signals with relatively low
elevation angles \cite{Milan_1997}. The 
signals potentially arriving from the back lobe were 
detected by their high elevation angle exceeding the threshold level
determined by the method \cite{Berngardt_2025b} over the years of
low and high solar activity. These threshold angles are determined
based on the raytracing results for measured signals
- their altitudes in case of main beam should be within the range
of ionospheric scattering altitudes (usually less than 350 km).
Due to raytracing and actual radar measurements 
 depends on geographical and
geophysical factors, the threshold level depends on the radar and year. 
The calculated threshold levels for given radars and years are shown in
the Table \ref{tab:thres}.

\begin{table}
\centering
\caption{Threshold elevation angles for separating signals into main and back
lobes (for years of low and high solar activity, and the maximum value
between them), degrees}
\begin{tabular}{|c|c|c|c|}
\hline 
\textbf{Radar} & \textbf{Low activity year} & \textbf{High activity year} & \textbf{Max threshold}\tabularnewline
\hline 
\hline 
CVE & 28 & 33 & 33\tabularnewline
\hline 
CVW & 28 & 35 & 35\tabularnewline
\hline 
EKB & 29 & 43 & 43\tabularnewline
\hline 
HOK & 29 & 35 & 35\tabularnewline
\hline 
KAP & 21 & 33 & 33\tabularnewline
\hline 
MAGW & 35 & 56 & 56\tabularnewline
\hline 
MCM & 14 & 19 & 19\tabularnewline
\hline 
PGR & 26 & 34 & 34\tabularnewline
\hline 
PYK & 22 & 28 & 28\tabularnewline
\hline 
SAS & 34 & 33 & 34\tabularnewline
\hline 
SPS & 34 & 17 & 34\tabularnewline
\hline 
STO & 27 & 31 & 31\tabularnewline
\hline 
\end{tabular}
\label{tab:thres}
\end{table}

For further training of classifier, the median value of the threshold
elevation angle over all maximums, 34 degrees, was chosen as the threshold
elevation angle for all the radars. The average value differs little from
it and is 34.6 degrees. Signals coming from elevation angles below
the threshold were considered to come from the main lobe of the antenna
pattern and were used for further analysis. The signals above thershold level
were considered to come from the back lobe of the antenna pattern
and were excluded from the model training and analysis.
We have not used standard threshold level calculated from antenna spacing discussed in \cite{Milan_1997},
because it works not good enough at ranges less 400km, and excludes too much actual main beam signals, 
especially meteor trail scattering used for the radar calibration.

\subsection{Pre-training and estimating the number of latent classes in the data}

The important problem of data analysis is determining the number of
latent (hidden) classes in the data - how many different types of
signals the radars actually observe. Traditionally, this number and
types of signals are postulated by the researcher before building
a classifier, and only the parameters of such a classifier are determined
during training \cite{Ponomarenko_2007,Blanchard_2009,Ribeiro_2011,Lavygin_2019,Kunduri_2022,KONG_2024}.
This approach is typical for supervised learning. In the data-driven
approach we propose that the number of hidden classes in the data is an
unknown parameter to be found from the measured data, and interpretation
of these classes will be carried out after the classifier is
trained and the data is separated into classes by it.

The search of a classifier for an unknown number of latent classes
was carried out using the methodology described in details in \cite{Berngardt_2025}:
at the first stage of training, the data of each experiment were clustered
using the GM+BIC method (Gaussian Mixture as a clusterer, the optimal
number of clusters corresponds to the minimum of Bayesian information
criterion). The applicability of such a technique was discussed in
detail in \cite{Berngardt_2025} - it gives a slightly overestimated
number of clusters, but leads to well-interpretable results after
the final training of the classifier. The data obtained
with high spectral resolution (i.e. using 16-pulse sequences at
EKB and MAGW radars instead of standard 7- or 8-pulse SuperDARN sequences)
were augmented by spectral width (their spectral widths were increased accordingly). 
At the second stage of training,
in accordance with \cite{Berngardt_2025}, a very wide two-layer classifier neural
network (300 and 140 neurons in the first and second
network layers of Encoder, N,M respectively in Fig.\ref{fig:architect}C)
with absolute activation functions was trained. We train it at the
data labeled at the first stage. At the third stage of training, according
to the method \cite{Berngardt2024a}, we find minimum sufficient number
of neurons  in the fully connected layers of the classifier network (i.e. optimize  N,M in Fig.\ref{fig:architect}C) 
to reach the forecast quality found at second stage. 
The minimum number
of neurons in output classification layer (M in Fig.\ref{fig:architect}C) corresponds 
to expected minimum
number of latent classes in the data. At the fourth stage of training,
the classifier was retrained again over the whole original data labeled
by clusterer, but with the found minimum number of neurons in each layer
of classifier.

The maximum number of clusters found by GM+BIC clusterer in each experiment
was 52. The above first three stages of training were conducted
and the minimum number of neurons in the classifier was found. An
obvious initial assumption was the dependence of the number of latent
classes in the measurements on the level of solar activity. Therefore, the classifier
was initially trained in the three variants: by using only the data over the years of
high solar activity, by using only the data over the years of low solar
activity, and by using the data over all years. 

\begin{figure}
  \centering
  \includegraphics[scale=0.6]{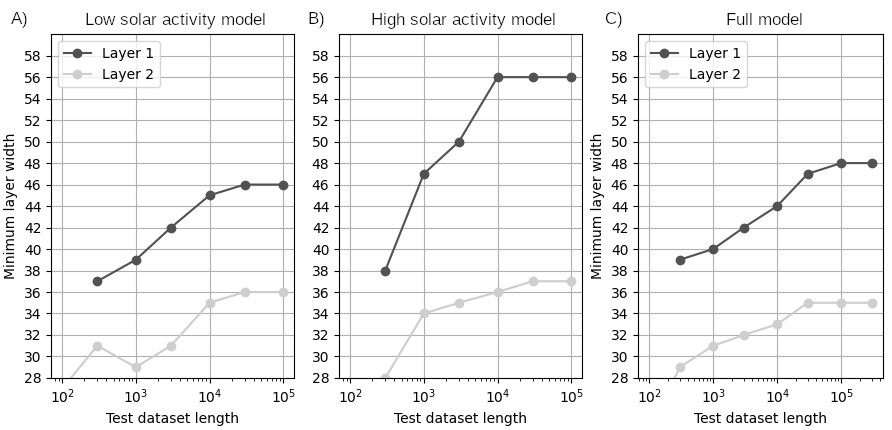}
  \caption{Minimum number of neurons in different layers of the classifier neural
network for years of low (A), high (B) solar activity, and for the
whole data set (C) depending on the size of the subset used for the
search. The size of Layer 2 is the expected number of latent classes
in the data, Layer 1 is the hidden layer size of the classifier.}
  \label{fig:OptimalLayerWidth}
\end{figure}

For each of the models at stage 3 we calculated the minimum
number of neurons in each classifier layer. The results are shown in Fig.\ref{fig:OptimalLayerWidth}.

The figure shows that the number of neurons in the output layer (Layer
2, M in Fig.\ref{fig:architect}C, the number of latent classes in the data) is almost the same for
the years of low and high solar activity and for the full dataset (36,
37 and 35 classes, respectively). This suggests that the number of
latent classes in the data is almost independent on the solar activity
level, and almost the same number of different types of scattered
signals is observed in all periods. It should be noted that this is
in good agreement with the value of 35 classes found earlier for 
EKB and MAGW radar data over the year of low solar activity
\cite{Berngardt_2025}. However, significant differences in the width
of the first layer (Layer 1,  N in Fig.\ref{fig:architect}C) indicate that it is much more difficult
to separate the data into classes at high solar activity
than at low activity (more neurons in the hidden layer of
the network are required to describe accurate boundaries between classes).
This is also consistent with experimental observations - during
high solar activity the signal interpretation is usually more difficult.

The figure shows that if the network is trained on the whole data
set, the parameters will have approximately the average value between
the years of low and high solar activity, which especially affects
the first layer, the size of which corresponds to the complexity of
classes separation. The number of latent classes
is practically the same.

\section{The final classifier}

\subsection{Choosing the optimal model}

It should be noted that a slight increase in the number of neurons
in the hidden layers of the neural network does not worsen the quality
of the approximation, but leads to an increase in the number of model
parameters. Therefore, in order to ensure good quality of
approximation in all solar activity conditions for a network trained
at all data (low and high solar activity), it is convenient to choose
56 neurons in the first layer and 37 neurons in the second layer,
according to the years of high solar activity. 
It should be noted that the classification problem is non-convex
(has more than one local minimum), so the optimal model depends on
the initial state used to train the model (i.e. on pretrain stage).

To find the optimal model and the
optimal number of classes in the data, the final classifier was trained in the
following five variants:

- Model L (Low activity): pre-training (estimate of the initial approximation
of the neural network coefficients) at a small data set extracted from
the years of low solar activity, followed by full training at the
entire data set of low solar activity. The optimal classifier size
corresponds to the years of low solar activity - 36 classes and 46 neurons
in the hidden layer (Fig.\ref{fig:OptimalLayerWidth}A);

- Model H (High activity): pre-training at a small data set extracted
from years of high solar activity, followed by full training at the
entire data set of high solar activity. The optimal size corresponds to the
years of high solar activity - 37 classes and 56 neurons in the
hidden layer (Fig.\ref{fig:OptimalLayerWidth}B);

- Full model A: pretraining at the data set extracted from the full
data set, followed by full training at whole data set. The optimal size
corresponds to the full dataset - 35 classes and 48 neurons in the hidden
layer (Fig.\ref{fig:OptimalLayerWidth}C);

- Full model B: pretraining at the small dataset extracted from the
years of high solar activity followed by full training at whole data
set. The optimal size corresponds to the years of high solar activity
- 37 classes and 56 neurons in the hidden layer (Fig.\ref{fig:OptimalLayerWidth}B);

- Full model C: pretraining at the dataset extracted from the whole
dataset, followed by full training at whole data set. The optimal
size corresponds to the high solar activity years - 37 classes and 56 neurons
in the hidden layer (Fig.\ref{fig:OptimalLayerWidth}B).

For all the models, training was performed by cross-validation over
three folds (three almost independently trained variants of each model
were obtained). Pretraining
allows choosing a good initial approximation and includes, at the
first, training the model (encoder + decoder) at 1000 random
experiments from the pretraining dataset, and at the second,
training the decoder at the entire training dataset without additional
training of the encoder. Pretraining is made in the following way to avoid
permutation of class numbers (this is an effect typical for hidden
layers of neural networks): during pretraining only one
variant has randomly inited network coefficients before training during cross-validation,
the other two variants use its trained coefficients as their initial
approximation before pretraining. The results for 5 network variants
for 3 cross-validated versions of each model are shown in the Table
\ref{tab:Cmp5networks}. Quality metric caluclation (area under the
precision-recall curve, AUC-RP, informative for complex data with
strong class imbalance) was calculated for each of the models at the corresponding
validation part of the dataset; at the test data set, the results
were close to them with error about $10^{-3}$.

Although statistically AUC-RP is approximately the same in all five
cases (in terms of its confidence interval), the last network variant
looks more acceptable - although it is slightly worse than
the model trained on years of high solar activity (model H), but its
quality was calculated at the whole dataset and it was trained at
the whole dataset. It becomes clear when comapring Full model C with Full model B
(it is model H but trained at the whole dataset):  Full model B looks slightly worse
than the Full model C.

\begin{table}
\centering
\caption{Results of training possible classifier models on three folds}
\begin{tabular}{|c|c|c|c|c|c|c|c|}
\hline 
\multicolumn{4}{|c|}{\textbf{Train procedure}} & \multicolumn{4}{c|}{\textbf{AUC-RP}}\tabularnewline
\hline 
\hline 
\textbf{Model} & \textbf{Dim} & \textbf{Pretrain} & \textbf{Train/Test} & \textbf{@1} & \textbf{@2} & \textbf{@3} & \textbf{Mean}\tabularnewline
\hline 
L & 36x46 & low activity & low activity & 0.9048 & 0.9080 & 0.9031 & 0.9053\tabularnewline
\hline 
H & 37x56 & high activity & high activity & 0.9040 & 0.9122 & 0.9067 & 0.9076\tabularnewline
\hline 
Full A & 35x48 & all & all  & 0.8945 & 0.9082 & 0.9062 & 0.9030\tabularnewline
\hline 
Full B & 37x56 & high activity  & all  & 0.8981 & 0.9087 & 0.9064 & 0.9044\tabularnewline
\hline 
Full C & 37x56 & all & all & 0.9044 & 0.9127 & 0.9038 & 0.9069\tabularnewline
\hline 
\end{tabular}
\label{tab:Cmp5networks}
\end{table}

Therefore, bellow by the Full model we will mean the Full
model C: with sizes determined by years of high solar activity, but
pre-trained and trained at the whole data set. Qualitatively, such
a model allows separating the signal classes both during
years of high and low solar activity sufficiently well .

It should be noted that any version of the model trained at one of
the folds could be used for the classification, as well as the ensemble
model - a model that predicts a class only if all three model cross-validation
variants predict the same class. In the subsequent detailed analysis
more accurate ensemble version was used.

\subsection{Identifying well-distinguished classes}

An important issue is finding well- and poorly distinguishable classes
from these 37 detected classes. This corresponds to the following statement
of the problem: which classes are objectively separated well by any
good classification model, and which classes can be detected only by some classification
models? A possible answer to this question can be found by comparing
the predictions of the three models (L, H, and Full C models). If we have
signal classes that are unambiguously classified by any of these models,
then these classes are well distinguishable in the data. The remaining
detectable classes are subjective and depend significantly on the
model used or from the researcher.

\begin{figure}
  \centering
\includegraphics[scale=0.5]{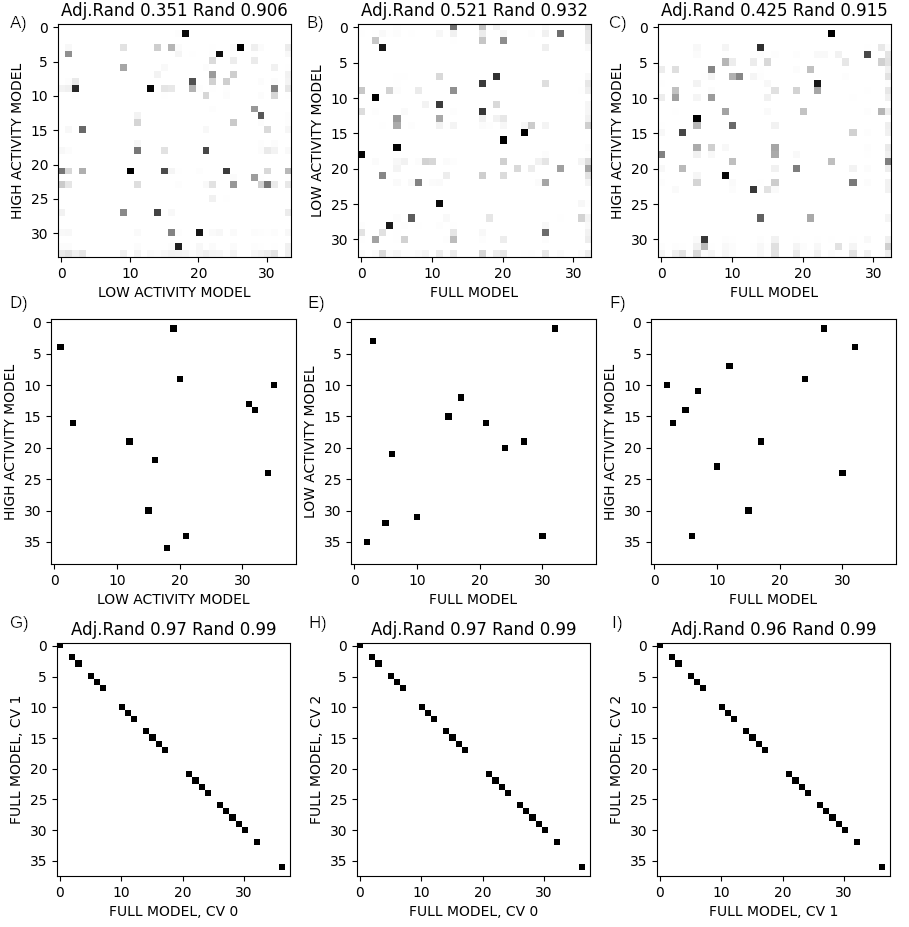}
\caption{Data classification comparsion for three models: correspondence between
classes defined by different algorithms - L, H, and Full. A-C) confusion
matrices for non-zero classes, D-F) matrices of best one-to-one correspondence
between classes detected by two models; G-I) matrix of best on-to-one
correspondence between variants of the Full model in the ensemble
(during cross-validation). The title of the plot indicates the value
of the Rand index and the Adjusted Rand index.}
\label{fig:RandCmp}
\end{figure}

To find well distinguishable classes, we compare how the three models
(L, H, and Full C) classify our data. Let us compare the results
between any two models. To do this, let us determine how likely
each class identified by some model (Model 1) is identified by another
model (Model 2). We can construct a confusion matrix - the number
of correspondences between different signal classes identified by these different models. The results for the three
possible pairs of the models are shown in Fig.\ref{fig:RandCmp}A-C.
It is found that the Rand index (the percent of pairs
of identically clustered points) is about 91\% and higher, which indicates
that, in average, 91\% of signal pairs are combined into classes by
all these models in the same way. The Adjusted Rand index (the excess
of the percent of identically clustered pairs of signals compared to
random clustering) is also acceptable, although numerically lower
(but at least 35\%).

To find the well-separated signal classes between which the most probable
one-to-one correspondence can be established, we perform the following
procedure:

- using the confusion matrix for each class of Model 1, we find the
most probable class of Model 2.

- using the confusion matrix for each class of Model 2, we find the
most probable class of Model 1.

- if the correspondence of classes between the models is the same
in both cases, we will consider the two found classes to be the most
probable one-to-one correspondent, or well-distinguished.

The matrices of most likely one-to-one class corespondence between
models are shown in Fig.\ref{fig:RandCmp}D-F.

The figure shows that the data we analyze contains 12 well-distinguished classes during years
of low solar activity and 14 - during years of high solar activity. 
We will discuss the interpretation of the detected
classes below. The remaining classes are subjective ones because could depend
on the model used.

\subsection{Interpretation of classes}

\begin{figure}
  \centering
\includegraphics[scale=0.6]{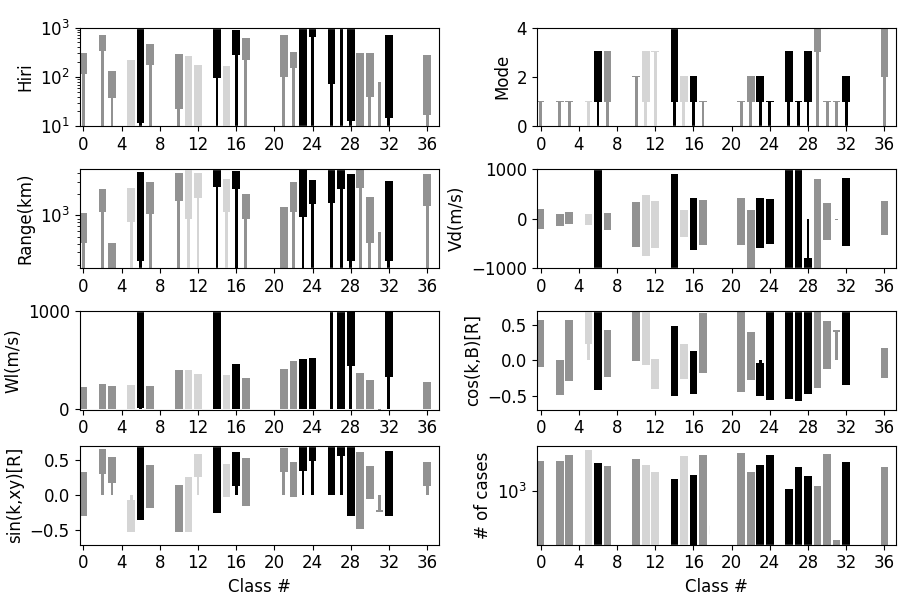}
\caption{Confidence intervals (95\%) of various signal parameters for each class. Black - uninterpreted signal types, 
gray - ionospheric scatter types, light gray - groundscatter types.}
\label{fig:classify}
\end{figure}

Unlike most standard approaches to classifying coherent scatter radar
data, in this data-driven approach we first classify the signals and
then interpret classes. To interpret the data, by analogy with \cite{Berngardt_2022a,Berngardt_2025},
statistics of various parameters for signals assigned to each signal
class are calculated. The parameters chosen are: the scattering height
according to the IRI ray-tracing results (Hiri), the signal hops
(Mode, the number of scattering/reflections from the underlying layers),
the radar range to the scatterer (Range), the Doppler velocity measured (Vd), 
the spectral width of the received signal in Doppler
velocity units (Wl), the cosine of the aspect angle of the radio wave
vector with the Earth's magnetic field at the calculated scattering
point (cos(k,B)), and the elevation angle of the trajectory at the
calculated scattering point (sin(k,xy)). In Fig.\ref{fig:classify}
for each class the 95\% confidence interval for each of these parameters
is shown. From the figure it is possible to make a qualitative interpretation
the main groups of classes: the signals, scattered from the earth's surface;
the signals, scattered from the ionosphere; and the signals, which are difficult
to interpret.

The group of classes of the signals scattered from the earth's surface
(marked by light gray color in Fig.\ref{fig:classify}) includes 4 classes:
5, 11, 12, 15. The main feature of these signals is their low scattering
height (below 100km), relatively low drift velocity and spectral width. Additional
information about the possible nature of scattering can be provided
by the mode - the number of reflections from the underlying surface.

The group of signal classes with complicated interpretation (marked by
black color in Fig.\ref{fig:classify}) includes 9 classes: 6, 14, 16,
23, 24, 26, 27, 28, 32. Their interpretation is complicated due to their
very high altitudes, unusual for the ionospheric scatter, and a large
spread of velocities or spectral widths. In the future, these classes
require more careful analysis and interpretation.

The group of classes of signals scattered from the ionosphere includes all other
signals (marked by gray color in Fig.\ref{fig:classify}). These
are 12 classes: 0, 2, 3, 7, 10, 17, 21, 22, 29, 30, 31, 36.

It should be noted that 12 classes are practically not observed in
the experiments (classes 1, 4, 8, 9, 13, 18-20, 25, 33-35), which
is due either they are very rare, or that they are
mathematically "balancing" classes necessary to provide the best fit between
the classification and clustering results at the training stage 
(the clustering could differ from actual physical classes separation and therefore is not optimal). This
could be also a sign that the absolute activation function we used
is not optimal, and replacing it with another activation function
may help to simplify the classifier and reduce the number of latent classes
without losing the classification quality.

According to the analysis, there are total 25 frequently observed classes in the data,
and 16 of them are interpretable classes, which are shown in Table
\ref{tab:interpret} together with another, poorly interpretable classes.
The 14 well-distinguished classes(Fig.\ref{fig:RandCmp}D-F) are highlighted in bold.

\begin{table}
\centering
\caption{Preliminary interpretation of some classes. Classes in bold 
are well-distinguished classes according to the models L, H,
and Full. }
%%\begin{tabular}{|c|>{\raggedright}p{4cm}|>{\raggedright}p{6cm}|}
\begin{tabular}{|c|c|c|}
\hline 
\#Full & Interpretation & Comments\tabularnewline
\hline 
\hline 
0 & E/F 0.5 hop & low V,W \tabularnewline
\hline 
\textbf{2} & \textbf{F 0.5 hop } & \textbf{low V,W}\tabularnewline
\hline 
\textbf{3} & \textbf{Meteor/near-range E 0.5 hop} & \textbf{low V,W}\tabularnewline
\hline 
\textbf{5} & \textbf{GS, 1hop} & \textbf{low V,W}\tabularnewline
\hline 
\textbf{6} & \textbf{Uninterpreted} & \textbf{too high H,V,W}\tabularnewline
\hline 
\textbf{7} & \textbf{F, 1-3hop} & \textbf{low V,W, aspect}\tabularnewline
\hline 
\textbf{10} & \textbf{E, 1.5hop} & \textbf{intermediate V,W}\tabularnewline
\hline 
11 & GS, 1-3hop & high V, intermediate W\tabularnewline
\hline 
\textbf{12} & \textbf{GS, 3 hop } & \textbf{high V, intermediate W}\tabularnewline
\hline 
\textbf{15} & \textbf{GS, 2 hop} & \textbf{low V,W}\tabularnewline
\hline 
\textbf{17} & \textbf{F 0.5 hop} & \textbf{intermediate V, low W}\tabularnewline
\hline 
\textbf{21} & \textbf{near-range F 0.5hop} & \textbf{intermediate V, low W}\tabularnewline
\hline 
22 & E/F 0.5hop & high V,W, aspect\tabularnewline
\hline 
\textbf{24} & \textbf{Uninterpreted} & \textbf{too high H}\tabularnewline
\hline 
\textbf{27} & \textbf{Uninterpreted} & \textbf{too high H,V,W}\tabularnewline
\hline 
29 & E, 3.5hop & high V, intermediate W\tabularnewline
\hline 
\textbf{30} & \textbf{E/F 0.5hop} & \textbf{intermediate V, low W}\tabularnewline
\hline 
31 &  \textbf{Rare class} & too few observations\tabularnewline
\hline 
\textbf{32} & \textbf{Uninterpreted class} & \textbf{too high H}\tabularnewline
\hline 
36 & E/F 2-4hop & low V,W, aspect\tabularnewline
\hline 
\end{tabular}
\label{tab:interpret}
\end{table}

Of these classes, 8 cases (2, 3, 5, 10, 15, 17, 21, 30) correspond
to scattering at ionospheric altitudes and below with velocities and
spectral widths less than 1 km/s, so they can be interpreted from
a radiophysical point of view. At high solar activity level, the are
two more classes exist, which are
difficult to distinguish at a low level of solar activity. These are
classes 7 and 12. 
We can interpret the 10 classes from a physical point of view as follows:
\begin{itemize}
\item Groundscatter at 1st hop (class 5);
\item Groundscatter at 2nd hop (class 15);
\item Groundscatter at 3rd hop (class 12, clearly distinguished during high solar activity);
\item Meteor/near-range E scatter at 0.5 hop (class 3); 
\item near-range F-layer scatter at 0.5 hop (class 21);
\item E or F layer scatter at 0.5 hop (class 30). 
\item F layer scatter at 0.5 hop with low V and W (class 2); 
\item F layer scatter at 0.5 hop, with intermediate V and low W (class 17);
\item E layer scatter at 1.5 hop with intermediate V and W (class 10);
\item F layert scatter at hops 1-3 with low V and W (class 7, clearly distinguished during
high solar activity);
\end{itemize}

Four classes (6,24,27,32) are poorly interpretable.

\subsection{Importance of different parameters for classification}

An important issue discussed today is the question of which
 parameters
need to be taken into account for accurate classification of the radar data, 
in particular the need to take into account
the elevation angle\cite{Ponomarenko2023}. Within the framework of
the proposed model, this task is reduced to the taks well-known in machine
learning - Feature Importance: determining the parameters that
significantly affect the result of the model. One of the feature importance algorithms 
was used in a similar task in \cite{KONG_2024}.

One of the universal methods that allows this to be done is the permutation
feature importance method \cite{PermutationFeatureImportance,HUANG2020},
in which the importance of an input parameter for a forecast is judged
by the change of the forecast quality when the values of this
parameter in the data set are randomly permuted.

\begin{figure}
  \centering
\includegraphics[scale=0.7]{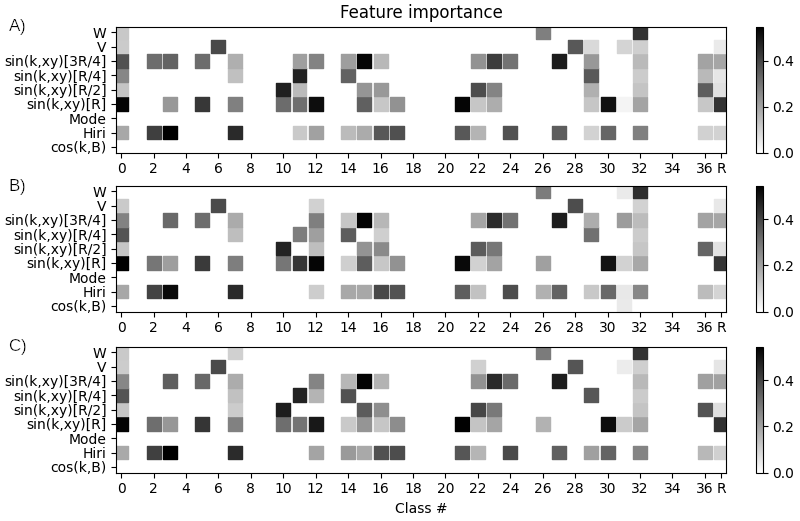}
\caption{Feature importance for different network variants trained on different
cross-validation folds (A,B,C) and for different classes. Column R
is the importance for the whole classification.}
\label{fig:feature_imp}
\end{figure}

Fig.\ref{fig:feature_imp} shows the importance of the different input
parameters for identifying the different classes according to this
method described in \cite{Berngardt_2025}, with a higher value corresponding
to more important parameter. The importance of the final classification
is also shown (column 'R'). The results are shown for each network
variant obtained during cross-validation train process. Empty cells 
correspond to minimally important parameters.

From Fig.\ref{fig:feature_imp} it is evident that the most important
parameters for classification are: the elevation angle of the radio
wave propagation trajectory at the scattering point and at 3/4 of
the path length, the ray-traced scattering height, and also nearly
equally the angle with the Earth's magnetic field at the scattering
point and the elevation angle in the middle of the signal propagation
trajectory. The least important parameters are the signal propagation mode
and the spectral width of the received signal. Thus, the most important
parameters for classifying scattered signals are the shape of the
radio signal propagation trajectory at its last  half and the scattering height. These
parameters cannot be measured directly by the radar and require simulation
of the radio wave propagation process. To do this, it is necessary
to know the sounding frequency, the three-dimensional antenna pattern,
the radiowave elevation angle and azimuth, and the
three-dimensional structure of the ionospheric refractive index. It
is obvious that in complex situations, when the propagation trajectory
is difficult to predict (there are no reliable measurements of the
elevation angle of the received radio wave or there is no sufficiently
accurate model of the ionosphere), this method will produce significant
errors. This explains the wide use of simpler classification methods
based on velocity and spectral width \cite{Ponomarenko_2007,Blanchard_2009}
at high-latitude radars. This could be seen from Fig.\ref{fig:feature_imp}: 
the spectral width is important for classification only for 0th,7th,26th,31st,32nd classes - 
the most difficult classes for interpretation from raytracing 
point of view (Fig.\ref{fig:RandCmp}F, Fig.\ref{fig:classify}, Table \ref{tab:interpret}).

For the radars with calibrated elevation angle measurements, the simulation
of the radio wave propagation trajectory with IRI model shows that the
classification method we developed can be successfully applied.

Unexpected results are weak dependence of the classifier quality
on the spectral width and on the trajectory angle with the magnetic
field. The first result can be explained by the presence of trajectory
information, which provide a better estimate of the scattering heights than
taking into account the spectral width. The second result can be caused
by errors in the ionospheric model and trajectory calculations, which
significantly complicate the verification of the conditions of aspect
sensitivity with the magnetic field.

The final classifier model thus has the following form. The output
(logit) of the model m for each of the classes k is:

\begin{equation}
y_{km}=\left|\sum_{j=1}^{56}C_{kjm} \left(\left| \sum_{i=1}^{9}A_{ijm}x_{i}+B_{jm} \right|\right)+D_{km}\right|
\label{eq:Ykm}
\end{equation}

m=1..3 is the model identifier in the ensemble, k=0..36 is the
class number, and $x_{i}$ are nine input parameters of the model
obtained from the measurements and propagation trajectory simulation.
Matrices A, B, C, D are model parameters found during model training.

The maximum probable class $K_m$ for a given signal determined by model
m is the class corresponding to the maximum logit $y_{km}$:

\begin{equation}
K_{m}=argmax_{k\in[0,36]}(y_{km})
\label{eq:Km}
\end{equation}

The signal class should be identified similarly by all three models in the ensemble
(or the signal is idintified as a new class 37, corresponding to uncertainly classified signals):

\begin{equation}
K=\left\{ \begin{array}{c}
K_{0},ifK_{0}=K_{1}=K_{2}\\
else\,37
\end{array}\right.
\label{eq:K}
\end{equation}

Fig.\ref{fig:RandCmp}G-I shows a comparison of the signal identification by all three
network variants in terms of confusion matrix, used before. 
It is evident that in most cases these 3 models provide almost identical
classification (Rand index $>$ 0.99, Adjusted Rand index $>$ 0.96). This
means that, if necessary, to simplify or speed up the classification, 
it is possible to use only one of the three models
trained in the cross-validation process, for example $y_{k0}$ (\ref{eq:Km}),
without using ensemble processing (\ref{eq:K}). Thus, the sufficient
model has the form (\ref{eq:Ykm},\ref{eq:Km}) and has 2669 coefficients.

To simplify the model one could take into account only 
25 of 37 classes actually
observed in the experiment (Fig.\ref{fig:classify}H),
rare classes can be ignored in the calculations and the number
of coefficients in the model can be reduced downto 1985.

The final trained ensemble version of the classifier model is available
at https://github.com/berng/SECIRA\_SD\_Classifier/.

\section{Discussion}

The dependence of different scattering types on geographical location and geophysical conditions is 
well known experimental fact.
Let us estimate the dependence of observation occurrence of different
classes at different radars depending on geophysical conditions. To
do this, based on the data of each radar (the test part of the dataset
that was not used to train the classifier), we calculate the
frequency (probability) of observations of each class. Calculating the obseravtion
frequency and, in general, a kind of normalizing the number of observations
is necessary due to radars can change their operating regimes, scanning
regimes and temporal resolution. 
Therefore the absolute number of observations
of a signal class would not be an indicative characteristic.

\begin{figure}
  \centering
\includegraphics[scale=0.5]{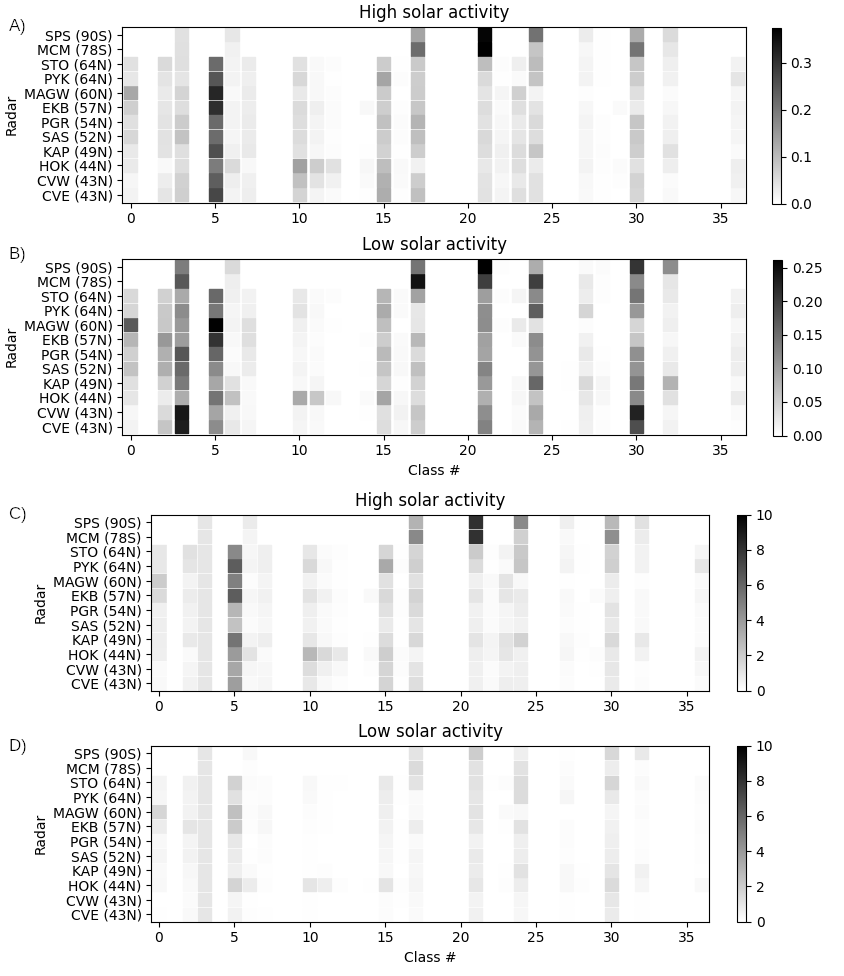}
\caption{Occurrence of different classes depending on the radar latitude and
solar activity level. A,B) - normalized to the total number of observations
at each radar, C,D) - normalized to the number of observations of
class 3 at each radar.}
\label{fig:rad_class_stats}
\end{figure}

Fig.\ref{fig:rad_class_stats}A-B shows the probability of observing
different classes by different radars depending on the level of solar
activity and the radar latitude. It is evident from the figure that
this dependence is significant. One can see the
decrease of the occurrence frequency of class 3 signals (meteor trail
scattering and near-range echo from the lower layers of the ionosphere)
and classes 21, 24, 30 (near-range 0.5 hop scatter from  F-layer; Uninterpreted
class; 0.5 hop scatter from E/F layers with medium V and low W, respectively).
Apparently, the decrease of the occurence frequency of class 3 may
be unjustified in most cases (meteor scattering should not depend
strongly on the level of solar activity, since it is more likely to
be associated with astronomical features of meteor showers than with
solar activity, and the radiowave absorption in the lower part of
the ionosphere is relatively small due to high elevation angles
to the meteor trails at short ranges). Therefore, it makes sense to
consider not the occurrence frequency of different echo types at each
radar, but their relation to the class with a nearly regular
occurrence, for example with meteor trail echo at the corresponding
radar.

Fig.\ref{fig:rad_class_stats}C-D shows the number of observations
of different classes relative to the number of observations of class
3 at the radar depending on the solar activity level
and the radar geographical latitude. These figures allow a fairly
convincing interpretation of the results from a physical point of
view.

It is obvious from the figure that the strongest effect is observed
in the increase in the class 5 signals - scattering from the earth's
surface at the first hop (1st hop groundscatter). This is easily
explained by the growth of the electron density during years of high
solar activity, and as a consequence - an increase of the signals
number propagating with reflection from the ionosphere, including
scattered from the earth's surface.

The increase of number of signals of the ionospheric scattering classes
(7,10,17,22,36), scattering from the earth's surface at the second
hop (15) and uninterpreted signals (6,23) also confirms the well-known
experimental fact that under high solar activity conditions the number
of scattered signals of various types increases: the number of signals
scattered from the ionosphere increases with the growth of ionospheric
small-scale irregularities, the number of signals scattered at the
second hop from the earth's surface increases with electron density
growth, and the number of uninterpreted signals increase due to the
growth of the amplitude of various large-scale ionospheric irregularities,
disturbing raytracing results.

A special group of radars is polar radars in south hemisphere - MCM
and SPS. They observe mainly only 7 classes of the signals. These
are 4 classes associated with scattering in the ionosphere: 3 (meteor
trail scattering), 17 (0.5hop F-layer scattering), 21 (0.5 hop near-range
F-layer scattering), 30 (0.5hop E/F-layer scattering), and three uninterpreted
classes: 6, 24, 32. These radars have not observed any groundscatter
signals. This is related with the ice covering nearby territory and preventing 
reflection back at off-perpendicular angles \cite{Ponomarenko_2010}.

Also interesting the group of three lowest-latitude radars: CVE,
CVW, and HOK. At high levels of solar activity they observe signal classes
that are less frequently observed by other radars - 11 and 12. The
classes are unusual types of scattering from the earth's surface -
with high velocities (class 11) and with three hops (class 12). The
appearance of class 12 can be explained by the following. For relatively
low-latitude radars, even a very long propagation trajectory can lie
below the polar oval - an area of strong absorption, so the signals
with a high number of hops are not affected by strong attenuation,
in opposite to higher latitude radars. The appearance of class 11
has not yet been explained, and may be associated, among other things,
with their incorrect interpretation as scattering from the earth's
surface (for example, this could be scattering from the E-layer at
a 1.5 hop).

\begin{figure}
  \centering
\includegraphics[scale=0.5]{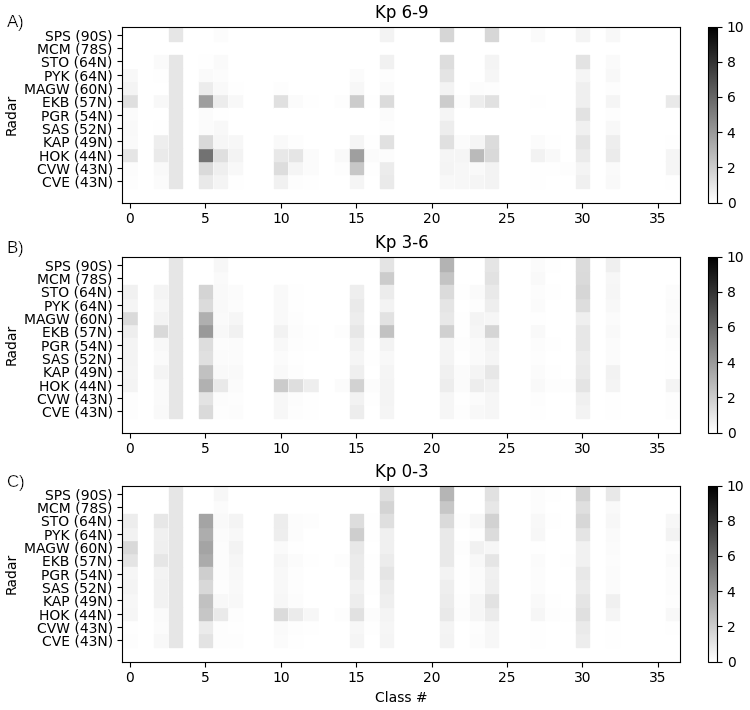}
\caption{Occurrence of different classes on the radars depending on the radar
latitude and the level of geomagnetic activity, normalized to the
number of observations of class 3 (meteors/E-layer near-range echo)
at each radar. There were no data on the MCM radar during the studied
periods with Kp 6-9.}
\label{fig:rad_class_kp}
\end{figure}

The dependence
of the occurrences of different classes normalized to observations
of class 3 on geomagnetic activity is shown in Fig.\ref{fig:rad_class_kp}.
It is evident from the figure that this dependence is significant.
Fig.\ref{fig:rad_class_kp} shows that occurance of most
classes, especially class 5 (1st hop groundscatter) decrease with
Kp growth. This is well known fact (radio blackout), associated with
the growth of ionospheric absorption and the decrease of electron
density during geomagnetic disturbances.

Also, Fig.\ref{fig:rad_class_kp} shows a significant increase of
some signal classes occurence at lower-latitude radars with increasing Kp: groundscatter of the
1st and 2nd hops (classes 5,15), scattering from the F-layer of the
1st-3rd hop (class 7), as well as an increase of occurence of uninterpreted
classes (6, 23, 24, 27, 28, 32). This is also explainable - during
significant geomagnetic disturbances, a radio blackout is observed
at most high-latitude radars, and only the lowest-latitude ones observe
scattered signals, including scattering in the F-layer, typical for
geomagnetic disturbances. During geomagnetic disturbances it is harder
to use IRI model for decribing actual ionosphere, so more signals
become uninterpretable.

\begin{figure}
\centering
\includegraphics[scale=0.45]{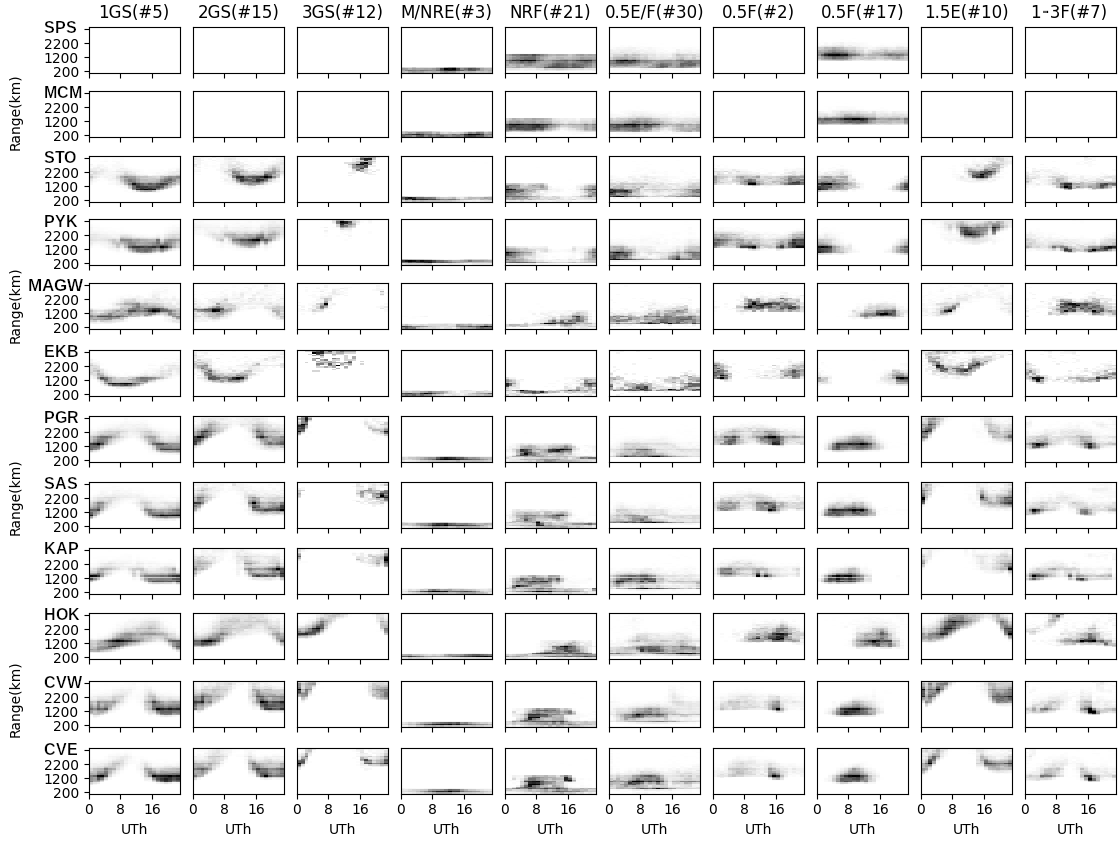}
\caption{Range-time occurrence of different classes at different radars during high solar activity years 
(the data not used for training).}
\label{fig:rad_class_rt}
\end{figure}

Fig.\ref{fig:rad_class_rt} shows the occurance of different classes at different radars
during high solar activity years at test data set, not used for training the classifier. 
One can see that found classes are correspndent between 
different radars and have explainable range-time dependence. Time offsets of same classes at different radars 
corresponds to local solar time difference at different longitudes.

\section{Conclusion}

The paper presents a method for automatic constructing a classifier
of processed data obtained by SuperDARN and SECIRA radars without
making additional assumptions. Method is based only on the radar data
, the results of automatic radiowave tracing in the 
IRI-2020 model ionosphere, and mathematical criteria for estimating
the quality of the models.

The final classifier is an ensemble model consisting of three network
variants trained on three random partitions of the available dataset
(12 radars of the SuperDARN and SECIRA networks, 1 year of low and
1 year of high solar activity for each radar) using the wrapped classifier
method, when clustering of the dataset is used to label it \cite{Berngardt_2022a}.
The number of detected classes in the data is 37, of which 25 are
frequently observed in the experiments, which is close
to the resullts obtained for EKB and MAGW radars in \cite{Berngardt_2025}.
The model has the form (\ref{eq:Ykm}-\ref{eq:K}). The number of
parameters for each network is 2669, which is significantly simpler
than the models \cite{Berngardt_2022a,Berngardt_2022b} and close
to the model \cite{Berngardt_2025}. When classifying radar data,
the model requires both calculated parameters of radio wave propagation
in the model ionosphere and parameters directly measured by the radar.
When calculating the elevation angle, the radars are calibrated using
scattering on meteor trails, assuming their 104km scattering altitude.

The analysis allowed us to identify 14 classes that were well-identified
by training differnet classification models. We can interpret 10 classes
of them from a physical point of view as 3 groundscatter classes and 
7 ionospheric scatter classes.
The remaining classes depend on the variant of the dataset used for
training, or are uninterpreted due to problems in calculating the
radiowave trajectory. Therefore in different cases (different models or researchers)
they can be grouped and interpreted differently, so require more detailed
study. It should be noted that the propagation model we use does not
distinguish between hops from the earth's surface and from underlying
layers, so when interpreting multi-hop propagation, the hops can be
associated not only with scattering from the earth's surface, but
also with interlayer waveguide propagation. Thus, in any geophysical
data analysis, it is looks useful to first pay attention to these
10 classes, since they will be most reliably distinguished by different
classification models or human researchers.

The analysis showed that from the parameters used for class identification,
the most important parameters are the calculated shape of the signal
propagation trajectory at its second half, the scattering height and the measured Doppler
velocity.

The average dynamics of observation of classes and their dependence
on the geographical latitude of radars at different levels of solar
and geomagnetic activity looks explainable and do not contradict known
patterns and mechanisms.

The final trained classifier model is available at https://github.com/berng/SECIRA\_SD\_Classifier/ .

\section*{Acknowledgments}

The work of OB and IL was supported by a grant from the Russian Science
Foundation \#24-22-00436, https://rscf.ru/project/24-22-00436/.

The authors acknowledge the use of SuperDARN data. SuperDARN is a
collection of radars funded by national scientific funding agencies
of Australia, Canada, China, France, Italy, Japan, Norway, South Africa,
United Kingdom and the United States of America.
SuperDARN data is available at https://www.frdr-dfdr.ca/repo/search?query=superdarn .
SECIRA data is available at http://sdrus.iszf.irk.ru/ekb/page\_example/simple . 

%% Bibliography
%% \bibliographystyle{abbrv}
%% \bibliography{REFS}

%%EDITED TILL HERE

\end{document}